\documentclass[runningheads]{llncs}

\usepackage{graphicx}
\usepackage[position=top]{subfig}
\usepackage{amsmath,amsfonts,amssymb,amscd}
\begin{document}
\title{Towards segmentation and spatial alignment of the human embryonic brain using deep learning for atlas-based registration}
\titlerunning{Towards segmentation and spatial alignment of the embryonic brain}
% If the paper title is too long for the running head, you can set
% an abbreviated paper title here
%
\author{Wietske A.P. Bastiaansen\inst{1,2}
\and Melek Rousian\inst{2} \and R\'egine P.M. Steegers-Theunissen \inst{2} \and Wiro J. Niessen \inst{1} \and Anton Koning \inst{3} \and Stefan Klein \inst{1}}

\authorrunning{W. Bastiaansen et al.}
%\authorrunning{** *********** ** ***}
% First names are abbreviated in the running head.
% If there are more than two authors, 'et al.' is used.

\institute{Departments of Radiology and Medical Informatics, Biomedical Imaging Group Rotterdam, Erasmus MC, Rotterdam, Netherlands \and
Department of Obstetrics and Gynecology, Erasmus MC, Rotterdam, Netherlands \and Department of Pathology, Erasmus MC, Rotterdam, Netherlands\\
\email{w.bastiaansen@erasmusmc.nl}}

\maketitle              % typeset the header of the contribution
\begin{abstract}
We propose an unsupervised deep learning method for atlas-based registration to achieve segmentation and spatial alignment of the embryonic brain in a single framework. Our approach consists of two sequential networks with a specifically designed loss function to address the challenges in 3D first trimester ultrasound. The first part learns the affine transformation and the second part learns the voxelwise nonrigid deformation between the target image and the atlas. We trained this network end-to-end and validated it against a ground truth on synthetic datasets designed to resemble the challenges present in 3D first trimester ultrasound. The method was tested on a dataset of human embryonic ultrasound volumes acquired at 9 weeks gestational age, which showed alignment of the brain in some cases and gave insight in open challenges for the proposed method. We conclude that our method is a promising approach towards fully automated spatial alignment and segmentation of embryonic brains in 3D ultrasound. 

\end{abstract}

\section{Introduction}
Ultrasound imaging is prominent in prenatal screening since it is noninvasive, real-time, safe, and has low cost compared to other imaging modalities \cite{Liu2019}. However the processing of ultrasound data is challenging due to low image quality, high variability of positions and orientations of the embryo, and the presence of the umbilical cord, placenta, and uterine wall. We propose a method to spatially align and segment the embryonic brain using atlas-based image registration in one unsupervised deep learning framework.  

Learning based spatial alignment and segmentation in prenatal ultrasound has been addressed before. In Namburete \cite{Namburete2018} a supervised multi-task approach was presented, which employed prior knowledge of the orientation of the head in the volume, annotated slices, and manual segmentations of the head and eye. Spatial alignment and segmentation was achieved on fetal US scans acquired at 22 till 30 weeks gestational age. Atlas-based registration was proposed by Kuklisova-Murgasova \cite{Kuklisova-Murgasova2013} where a MRI atlas and block matching was used to register ultrasound images of fetuses of 23 till 28 week gestational age. Finally Schmidt \cite{schmidt2017} proposed a CNN and deformable shape models to segment the abdomen in 3D fetal ultrasound. All these works focus on ultrasound data acquired during the second trimester or later and rely on manual annotations. Ground truth segmentations for our application were not available and are laborious to obtain, which motivated our unsupervised approach.

Developing methods for processing of ultrasound data acquired during the first trimester is of great clinical relevance, since the periconception period (14 weeks before till 10 weeks after conception) is of crucial importance for future health \cite{Steegers-Theunissen2016}. Therefore our method is developed for first trimester ultrasound.

Recently there has been quite some attention for unsupervised deep learning approaches for image registration, since these methods circumvent the need for manual annotations. Several methods were developed to learn dense nonrigid deformations under the assumption that the data is affinely registered \cite{Balakrishnan2018,Yang2017}. Employing multi-level or multi-stage methods, affine registration can also be included \cite{Hering2019,Hu2018,DeVos2019}. The framework presented here is based on the method presented in \cite{Balakrishnan2018} and follows the idea of \cite{Hering2019,Hu2018,DeVos2019} to dedicate part of the network to learn the affine transformation.

To the best of our knowledge this is the first work that addresses the development of a framework for the alignment and segmentation of the embryonic brain, captured by ultrasound during the first trimester, applying unsupervised deep learning methods for atlas-based registration. Segmentation and alignment are important preprocessing steps for any image analysis task, hence this method contributes to our ultimate goal: further improve precision medicine of human brain disorders from the earliest moment in life.

\section{Method}
Let $I$ and $A$ be two images defined in the $n$-$D$ spatial domains $\left (\Omega_{I}, \Omega_{A} \right) \in \mathbb{R}^n$, with $I$ the target image and $A$ the atlas. Both images contain single-channel grayscale data. Assume that $A$ is in standard orientation and the segmentation $S_{A}$ is available. Our aim is to find two deformations $\phi_a$ and $\phi_d$ such that:
\begin{equation} A(x)\approx I\left(\phi_a \circ \phi_d(x)\right) \quad \forall x \in \Omega_{A}, \end{equation}
where $\phi_a$ is an affine transformation and $\phi_d$ a voxelwise nonrigid deformation.

To obtain $\phi_a$ and $\phi_d$ a convolutional neural network (CNN) is used to model the function $g_{\theta}$: $(\phi_a,\phi_d)=g_{\theta}(I,A)$, with $\theta$ the network parameters. The affine transformation $\phi_a:=Tx$ is learned as a $m$-dimensional\footnote{For $n=2$, $m=6$ and for $n=3$, $m=12$} vector containing the coefficients of the affine transformation matrix $T\in\mathbb{R}^{(n+1) \times (n+1)}$. The voxelwise nonrigid deformation is defined as a displacement field $u(x)$ with $\phi_d:=x+u(x)$.

\begin{figure}[h]
    \centering
    \includegraphics[scale=0.4]{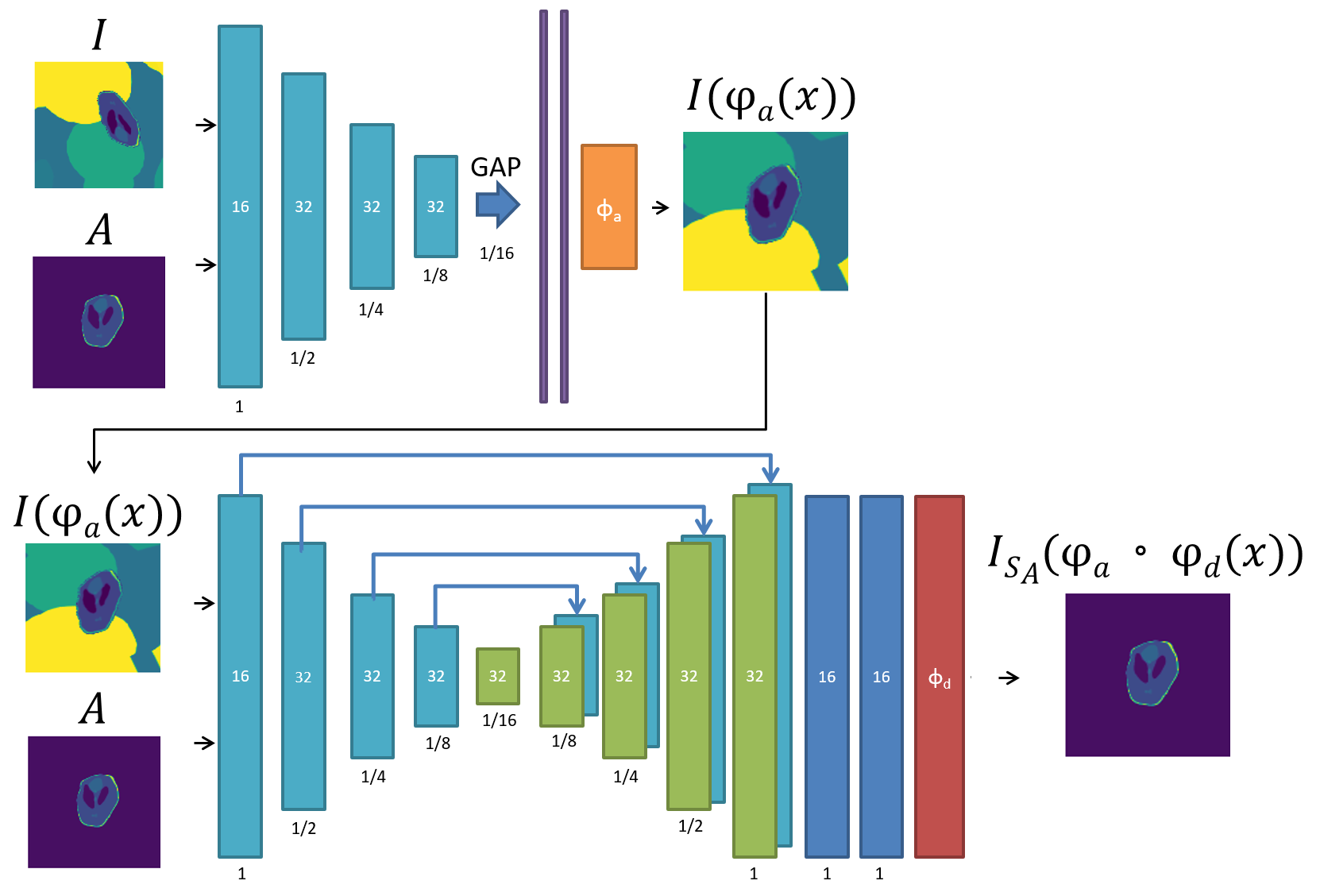}
    \caption{Architecture of our network. Light blue: convolutional layers with a stride of 2 (encoder). Green: convolutional layers with stride of 1, skip-connection, up-sampling layer (decoder). Purple: fully connected layers with 500 neurons and ReLU activation. Dark blue: convolutional layers at full resolution. Orange: $\phi_a$, red: $\phi_d$. All convolutional layers have a kernel size of 3 and have a LeakyReLU with parameter $0.2$. }
    \label{fig:network_gl}
\end{figure}

Figure \ref{fig:network_gl} provides an overview of our method. The input of the network is an image pair consisting of the atlas $A$ and target image $I$. The first part of the network outputs $\phi_a$ and the affine registered image $I(\phi_a(x))$. The input of the second part is the affinely registered image together with atlas $A$. The final output of the network consists of $\phi_a$, $\phi_d$, along with the registerd and segmented target image $I_{S_A}(\phi_a \circ \phi_d(x))=S_A(x)\cdot I(\phi_a \circ \phi_d(x))$ and the affinely registered image $I(\phi_a(x))$\footnote{Note that $I(\phi_a(x))$ is not segmented, since this is an intermediate result.}. 

Since this is an unsupervised method no ground truth deformations are used for training. The parameters $\theta$ are found by optimizing the loss function on the training set. The proposed loss function is described in the next section. After training, a new image $I$ can be given to the network together with the atlas to obtain the registration. 
\subsection{Network architecture}
 The target image $I$ and atlas $A$ are fed to the network as a two-channel image. The first part of the network consists of an encoder where the images are down-sampled, followed by a global average pooling layer. The global average pooling layer outputs one feature per feature map, which forces the network to encode position and orientation globally, and is followed by fully connected layers. The output layer consists of the entries of the affine transformation matrix $T$. The architecture of the second part of the network is the same as Voxelmorph \cite{Balakrishnan2018} and consists of an encoder and decoder and convolutional layers at full resolution. The output layer contains the dense displacement field $u(x)$. 

The method is implemented using Keras \cite{F.Cholletetall2015} with Tensorflow backend \cite{Tensorflow}. The ADAM optimizer is used with a learning rate of $10^{-4}$. Each training batch consist of one pair of volumes and by default we use 500 epochs.

\subsection{The loss function}
The loss function is defined as follows:
\begin{align}\begin{split}\label{eq:loss} \mathcal{L}(A,I,\phi_d,\phi_a)=&\mathcal{L}_{\text{sim}}\left[A,I\left(\phi_a \circ \phi_d (x)\right)\right] +\lambda_{\text{diffusion}} \mathcal{L}_{\text{diffusion}}\left[\phi_d \right]\\&  + \lambda_{\text{scaling}} \mathcal{L}_{\text{scaling}}\left[\phi_a\right]. \end{split}\end{align}
The first term promote intensity based similarity between the atlas and the deformed image, the second and third therm regularize $\phi_d$ and respectively $\phi_a$. Each term is discussed in detail below.

Since in 3D first trimester ultrasound there are other objects in the volumes besides the brain, the similarity terms are only calculated within the region of interest defined by segmentation of the atlas $S_A$. $\mathcal{L}_{\text{sim}}$ is chosen as either the mean squared error (MSE) or cross-correlation (CC). They are defined as follows:
\begin{align}
&\text{MSE}(A,Y)= \frac{1}{M} \sum_{p \in \Omega} W(p)\cdot \left (A(p)-Y(p) \right)^2\\
&\text{CC}(A,Y)= \notag\\
&\label{eq:cc}\frac{1}{M} \sum_{p \in \Omega} W(p)\cdot \frac{\left ( \sum_{p_i} [A(p_i)-\bar{A}(p)][Y_{S_A}(p_i)-\bar{Y}_{S_A}(p)]\right)^2}{\left ( \sum_{p_i} [A(p_i)-\bar{A}(p)]^2\right)\left(\sum_{p_i} [Y_{S_A}(p_i)-\bar{Y}_{S_A}(p)]^2\right)} ,\end{align}
where $M$ is the number of nonzero elements in $W$, unless stated otherwise $W=S_A$, the subscript $S_{A}$ indicates segmented, $\bar{A}$ and $\bar{Y}$ denote: $\bar{A}(p)=A(p)-\frac{1}{j^3} \sum_{p_i} A(p_i)$, where $p_i$ iterates over a $j^3$ volume around $p\in \Omega$ with $j=9$ as in \cite{Balakrishnan2018}.

Image registration is an ill-posed problem; therefore regularization is needed. $\phi_d$ is regularized by:
\begin{equation} \mathcal{L}_{\text{diffusion}}(u)= \frac{1}{M} \sum_{p \in \Omega} \|\nabla u(p)\|^2,\end{equation}
which penalizes local spatial variations in $\phi_d$ to promote smooth local deformations \cite{FischerBernd}. 

Initial experiments revealed that, when objects in the background of the target image are present, the affine transformation degenerate towards extreme compression or expansion. To prevent this, extreme zooming is penalized as regularization for $\phi_a$. The zooming factors must be extracted for $T(x)$. This is done using the Singular Value Decomposition (SVD) \cite{golub1971}, which states that any square matrix $T \in \mathbb{R}^{n\times n}$ can be decomposed in the following way:
\begin{equation} T=U\Sigma V^*,\end{equation}
where the diagonal matrix $\Sigma$ contains non-negative real singular values representing the zooming factors. The scaling loss is defined as:
\begin{equation}\label{eq:scaling}\mathcal{L}_{\text{scaling}}= \|\text{Diag}(\Sigma) - S \|_1.  \end{equation}
with $S$ an $n$-dimensional vector containing ones. 

For $\lambda_{\text{diffusion}}$ and $\lambda_{\text{scaling}}$ the optimal values must be chosen. This is addressed in the experiments.

\section{Data}
The following three datasets were used in the experiments.

\subsection{Synthetic 2D dataset 1}
To develop and validate our method against a ground truth, we created two synthetic 2D datasets. These synthetic datasets were created by affinely transforming and nonrigidly deforming the synthetic atlas. As synthetic atlas the Shepp-Logan phantom \cite{shepplogan} is used, which was nonrigidly deformed. The first dataset was created by first applying a random affine transformation $\bar{\phi}_a^{-1}$ on the atlas, followed by a nonrigid deformation $\bar{\phi}_d^{-1}$.

The coefficients for the affine transformation matrix $\bar{\phi}_a^{-1}(x):=T_{gt}^{-1}x$ were drawn as follows: translation coefficients $t_x$, $t_y$ $\in[0,40]$ pixels, rotation angle $\theta \in [0,360]$ degrees, anisotropic zooming factors $z_x$, $z_y$ $\in [0.5,1.5]$, and shear stress in the x direction $\theta_{s}$ $\in [0,30]$ degrees. The nonrigid deformation $\bar{\phi}_d^{-1}(x):=x+\alpha u_{gt}^{-1}(x)$ was generated using a normalized random displacement field $u_{gt}^{-1}(x)$, were $\alpha$ defines the magnitude of the displacement. The smoothness of $u_{gt}^{-1}(x)$ is controlled using $\sigma$, representing the standard deviation of the Gaussian, which was convolved with $u(x)$. We used $\alpha=40$, and $\sigma \in [3,7]$.

\subsection{Synthetic 2D dataset 2}
The second synthetic dataset was created in the same manner as the first, with additionally a background consisting of ellipses which have a random size and orientation. The ellipses are around, behind and adjacent to the synthetic atlas, to mimic the presence of the uterine wall around the embryo, and the body of the embryo attached to the head. Both datasets contain 3000 training, 100 validation and 100 test images.

\subsection{3D ultrasound data: Rotterdam Periconceptional cohort}
The Rotterdam Periconceptional Cohort (Predict study) is a large hospital-based cohort study embedded in tertiary patient care of the department of Obstetrics and Gynaecology, at the Erasmus MC, University Medical Center Rotterdam, the Netherlands. This prospective cohort  focuses on the relationships between periconceptional maternal and paternal health and fetal growth development, and underlying (epi)genetics \cite{Steegers-Theunissen2016}. 

Scans collected at 9 weeks gestational age were used as proof of concept for our method. The image chosen as atlas was put in standard orientation and had sufficient quality to segment the embryo and brain semi-automatically using Virtual Reality \cite{Rousian2018}.  There were 170 3D ultrasound scans available with sufficient quality, 140 are used for training and 30 for testing. All scans were padded with zeros and re-scaled to $64 \times 64 \times 64$ voxels to speed up training.

Since 140 scans is not sufficient for training, data augmentation was applied. When considering a 2D slice, the embryo is either visible in the coronal, saggital, or axial view. To keep this property during augmentation, first an axis was selected at random and a rotation was applied of either $90$, $180$ or $270$ degrees. Subsequently a random rotation on this axis was applied between $0$ and $30$ degrees followed by a translation $t_x, t_y, t_z \in[-15,15]$ and anisotropic zooming $z_x, z_y, z_z \in [0.9,1.3]$. Each volume was augmented 30 times and this resulted in 4340 images for training.

\section{Experiments}
To validate our method three experiments are performed. 
\begin{enumerate}
    \item Comparison with Voxelmorph \cite{Balakrishnan2018} on synthetic dataset 1 and $\mathcal{L}_{\text{sim}}= \text{MSE}$.
    Goal: evaluate influence of adding a dedicated part of the network for affine registration on images where the object of interest has a wide variation in position and orientation.
    \item Evaluation of hyperparameters in loss function Eq. \eqref{eq:loss} on synthetic dataset 2 and $\mathcal{L}_{\text{sim}}=\text{MSE}$. Goal: set $\lambda_{\text{diffusion}}$ and $\lambda_{\text{scaling}}$ in the presence of objects in the background.
    \item Testing method on 3D ultrasound data acquired at 9 weeks gestational age with $\mathcal{L}_{\text{sim}}=\text{CC}$ and different types of atlases as input for the network. $\mathcal{L}_{\text{sim}}=CC$ is used, since it is well known that the cross-correlation is more robust to intensity variations and noise
\end{enumerate}
The main difference between the synthetic data and ultrasound data is that for the synthetic data the atlas is the only object with a clear structure, while the ultrasound data is noisy and more structures similar to the embryonic brain are present, for example the body of the embryo. The body of the embryo is also a prominent round structured shape. To address this, in the third experiment the influence of using an atlas containing the whole embryo versus only the brain is evaluated. Using the atlas containing the whole embryo as input gives more information for alignment. However we aim at registering only the brain, since this is our region of interest and registering the whole embryo introduces new challenges due to movement and wide variation in position of the limbs. To focus on registration of the brain, $W(x)$ in Eq. \ref{eq:cc} is adjusted by assigning twice as much weight to the loss calculated in voxels that are part of the brain.

\subsection{Evaluation}
In the synthetic case the Target Registration Error (TRE) was calculated, which was defined as the mean Euclidean distance between $x_i \in \mathbb{R}^2$ for $i$ in the set of evaluation points:
\begin{equation}TRE\left[\bar{\phi}_a^{-1}, \bar{\phi}_d^{-1}, \phi_a, \phi_d \right]= \frac{1}{n}\sum_{i=1}^n \| \bar{\phi}_a^{-1} \circ \bar{\phi}_{d}^{-1} \circ \phi_a \circ \phi_d(x_i)-x_i \|, \end{equation}
where the evaluation points mark the boundary of the shape and important internal structures. The TRE is given in pixels.

In the case of real ultrasound data we visually asses the quality of alignment in the 30 test images. The following scoring is used: 0: fail, 1: correct orthogonal directions, 2: brain and atlas overlap, 3: alignment. Where score 1 indicates the network was able to detect the correct plane, score 2 indicates the network was able to map the brain to the atlas and 3 indicates successful alignment.

\section{Results}
\begin{figure}[h!]
    \centering
    \includegraphics[scale=0.35]{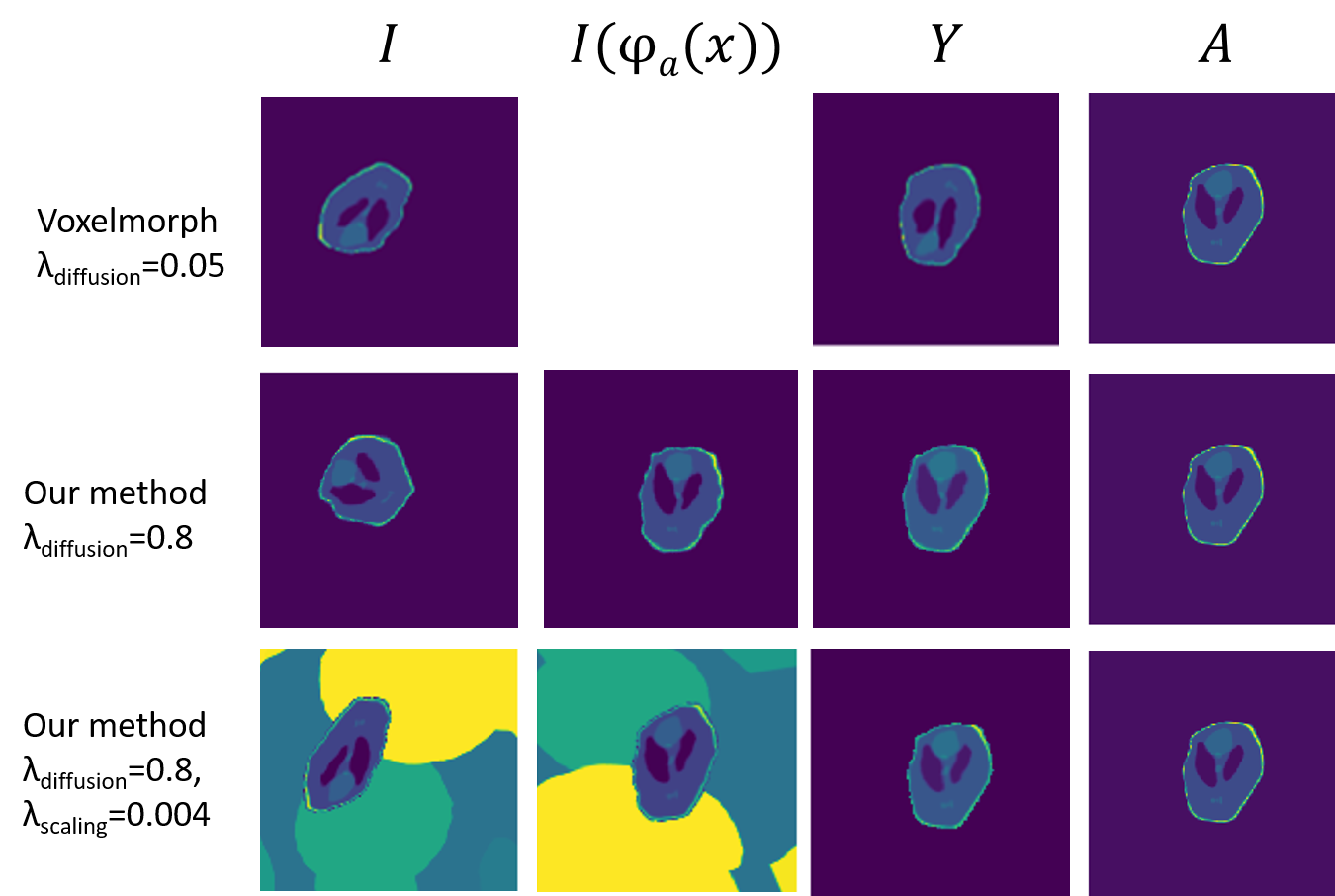}
    \caption{Visual result for experiment 1 and 2, $Y=I(\phi(x))$ in case of Voxelmorph architecture, $Y=I(\phi_a \circ\phi_d(x))$ for our method and $A$ the atlas.}
    \label{fig:visualexp2}
\end{figure}

In the first experiment we compared our method with Voxelmorph \cite{Balakrishnan2018} on the first synthetic dataset. The experiment was done for different values of $\lambda_{\text{diffusion}}$ with $\lambda_{\text{scaling}}=0$. Table \ref{tab:setlambdadiffusion} shows that with the architecture of Voxelmorph it was not possible to capture the global transformation needed. This is also illustrated by row one in Fig. \ref{fig:visualexp2}. Using our method a small TRE was achieved for both the train and validation set, see row 2 of Fig. \ref{fig:visualexp2} for an example. Setting $\lambda_{\text{diffusion}}=0.8$ gave a TRE of $2.71 \pm 1.67$ pixels on the test set, which is comparable to the result on the train and validation set. 
\begin{table}[h!]
    \centering
        \caption{Performance on first synthetic dataset using Voxelmorph \cite{Balakrishnan2018} and our method for different values of $\lambda_{\text{diffusion}}$. TRE is expressed in pixels, standard deviation between brackets.}
    \label{tab:setlambdadiffusion}
    \begin{tabular}{cccccccccc}
    \hline
     & \multicolumn{3}{c}{Voxelmorph} && \multicolumn{5}{c}{Our method}\\
    $\lambda_{\text{diffusion}}$ & Train && Validation && Train && Validation&&Test \\
     \hline
    0.05 & 34.27 (12.10) && 34.87 (11.35)&& 3.46 (6.86)&& 4.25 (8.35) && -\\ 
    0.2 & 34.15 (12.85) && 35.23 (12.24)&&2.71 (5.80) &&3.63 (7.25)&&-\\
    0.8 & 40.40 (12.67) && 42.12 (11.80)&& 2.20 (0.77) && 3.10 (1.78)&& 2.71 (1.67)\\
    3.2 &- &&- && 32.61 (34.07) && 35.60 (33.25)&&- \\ 
    \hline
    \end{tabular}
\end{table}
\newpage
In the second experiment we evaluated how to deal with objects in the background by penalizing extreme zooming. In Tab. \ref{tab:exp2background} one can find the results for $\lambda_{\text{diffusion}}=0.2$ and $\lambda_{\text{diffusion}}=0.8$ and for different values of $\lambda_{\text{scaling}}$. Setting $\lambda_{\text{scaling}}$ too high restricts the network to much, setting this value too low causes extreme scaling. The best result on the validation set was found for $\lambda_{\text{diffusion}}=0.8$ and $\lambda_{\text{scaling}}=0.004$, using this model to register the test set gave a TRE of $2.90 \pm 1.97$ pixels, which is again comparable to the result for the training and validation set. An example can be found in row three of Fig. \ref{fig:visualexp2}.

\begin{figure}[h!]
    \centering
    \includegraphics[scale=0.28]{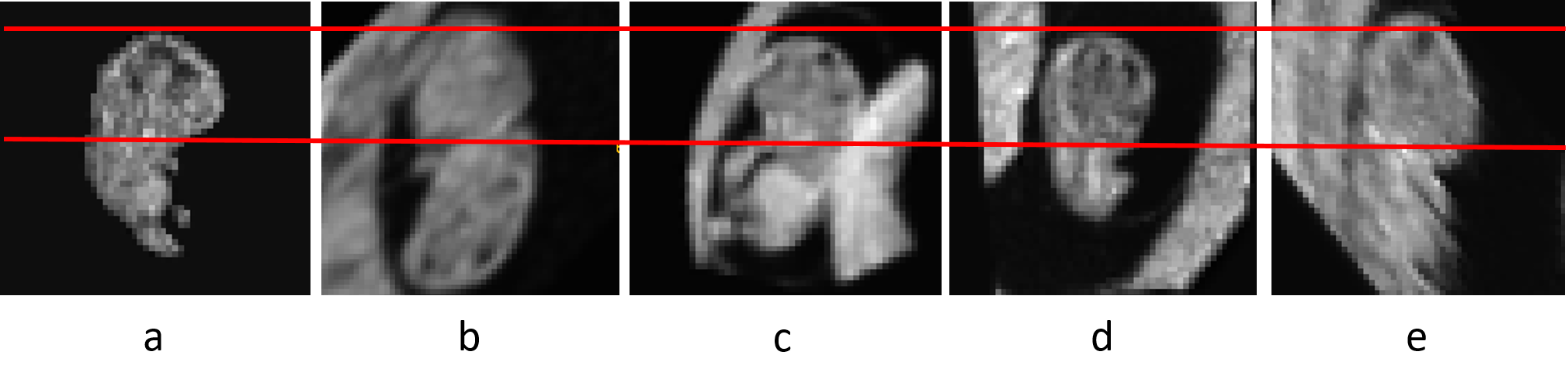}
    \caption{Same slice for: a) ultrasound atlas, b) example of image after alignment with score 1, c) example of image after alignment with score 2, d,e): example of successfully affine aligned images with score 3. Red line indicates correct boundaries of the brain after alignment.}
    \label{fig:3D}
\end{figure}

In the third experiment we evaluated our method on real ultrasound data, for different combinations of atlases as input to the two parts of the network. The results are shown in Tab. \eqref{tab:exp3}. Using the atlas of the whole embryo gives the best results, since the network has more information for alignment. Figure \ref{fig:3D} gives an impression of the resulting registrations. Note that the images that are marked as aligned are not perfectly registered, this is caused by the fact that the network still roughly misaligned most images and therefore voxelwise alignment is not learned.
\begin{table}[t!]
    \centering
        \caption{Target registration error for different hyperparameter settings of the loss function. TRE is expressed in pixels. The standard deviation is given between brackets.}
    \label{tab:exp2background}
    \begin{tabular}{cccccc}
    \hline
    $\lambda_{\text{diffusion}}$&$\lambda_{\text{scaling}}$ & Train & Validation & Test\\
    \hline
    0.2&0 & 4.02 (8.26)  & 5.43 (11.17) &-\\
    0.8&0 & 2.17 (3.64) & 2.74 (2.30) &-\\ 
    \hline
    0.2 &0.004  & 3.17 (3.08) & 3.26 (1.46) &- \\
    0.8 & 0.004 &2.36 (3.53) &2.45(3.53)& 2.90 (1.97) \\
    0.2 & 0.008 & 6.99 (10.26) &  6.25 (7.52)&- \\
    0.8 & 0.008 & 2.47 (3.35) & 2.53 (1.10)&- \\
    \hline
    \end{tabular}
\end{table}
\newpage
\begin{table}[h!]
    \centering
        \caption{Performance on ultrasound data for different type of atlas. Scoring: 0: fail, 1: correct orthogonal directions, 2: brain and atlas overlap, 3: alignment.}
    \label{tab:exp3}
    \begin{tabular}{ccccccccccc}
    \hline
   Part 1&& Part 2 && 0  &&1 && 2 && 3\\
    \hline
    Brain && Brain &&21 &&7 &&2 &&0 \\
    Embryo && Brain && 10&& 14 &&5 &&1\\
    Embryo && Embryo &&8&& 14 && 5   && 3     \\ 
    \hline
    \end{tabular}
\end{table}
\section{Conclusion}
 In this work we extended existing deep learning methods for image registration to developed an atlas-based registration method to align and segment the embryonic brain. Main extensions are the dedicated part of the network for affine registration and the loss function \eqref{eq:loss}. For validation, synthetic 2D datasets containing a ground truth were used. These experiments showed that our method can deal with the wide variation in position and orientation and with simple objects in the background. 
 
 The final experiment using real 3D ultrasound data acquired during the first trimester showed that our method is not robust enough to align and segment the embryonic brain. The importance of the atlas was evaluated and it turns out that using an atlas of the whole embryo improves results slightly, since it gives more information. This information is needed since the images are noisy, have artefacts and the embryonic brain is small (on average only $1\%$ of the volume).  Another drawback is that the ultrasound images were rescaled to one-fourth of the original size and during registration the image is resampled twice which makes the deformed image blurry and this has influence on the calculated loss function. The rescaling was done to speed up training.
 
Another way to speed up training, is to train in two stages. The second part of the network learning the voxelwise registration, can only learn useful features when the images are already roughly aligned. So training first the affine part of the network is more efficient, since from the start the second part can then learn useful features for voxelwise alignment. This will be explored in the future.

Finally, we aim to extend our method to be applicable to the entire first trimester, to enable spatio-temporal modeling of the embryonic brain. This extension can be made by training different networks for each period. Another natural extension is multi-atlas image segmentation \cite{IGLESIAS2015205}, both for networks trained within a certain period to get more robust results, or with a set of atlases covering the whole first trimester. 

\bibliography{references}{}
\bibliographystyle{splncs04}

\end{document}